\documentclass[titlepage]{article}
\usepackage[]{graphicx}
\usepackage[]{color}

\usepackage{amsmath,amsfonts,amssymb,amsthm,epsfig,epstopdf,titling,url,array}


\usepackage[section]{placeins}
\usepackage{setspace}
\theoremstyle{plain}

\theoremstyle{definition}
\newtheorem{defn}{Definition}[section]

\theoremstyle{remark}

\makeatletter
\def\maxwidth{ %
  \ifdim\Gin@nat@width>\linewidth
    \linewidth
  \else
    \Gin@nat@width
  \fi
}
\makeatother

\definecolor{fgcolor}{rgb}{0.345, 0.345, 0.345}

\usepackage{framed}
\makeatletter
 {\par\unskip\endMakeFramed%
 \at@end@of@kframe}
\makeatother

\definecolor{shadecolor}{rgb}{.97, .97, .97}
\definecolor{messagecolor}{rgb}{0, 0, 0}
\definecolor{warningcolor}{rgb}{1, 0, 1}
\definecolor{errorcolor}{rgb}{1, 0, 0}

\usepackage{alltt}
\usepackage{enumitem}
\usepackage{amsthm}
\usepackage{amsmath}
\usepackage{amssymb}
\usepackage{amsfonts}

\usepackage[style=authoryear,maxcitenames=2, doi=false]{biblatex}

\bibliography{references}

\usepackage[letterpaper, portrait, lmargin=1in, rmargin=1in,
bmargin = 1.35in, tmargin = 1.35in]{geometry}
\usepackage[english]{babel}
\usepackage{graphicx}
\usepackage{float}

\usepackage{caption}

\IfFileExists{upquote.sty}{\usepackage{upquote}}{}

\title{Canonical Least Favorable Submodels\\
A New TMLE Procedure for Multidimensional Parameters}
\author{Jonathan Levy}

\begin{document}

\begin{titlepage}

\maketitle
\begin{abstract}
This paper is a fundamental addition to the world of targeted maximum likelihood estimation (TMLE) \parencite{Laan:2006aa} (or likewise, targeted minimum loss estimation) for simultaneous estimation of multi-dimensional parameters of interest.  TMLE, as part of the targeted learning framework \parencite{Laan:2015ab}, offers a crucial step in constructing efficient plug-in estimators for nonparametric or semiparametric models.  The so-called targeting step of targeted learning, involves fluctuating the initial fit of the model in a way that maximally adjusts the plug-in estimate per change in the log likelihood \parencite{Laan:2015ab}.  Previously for multidimensional parameters of interest, iterative TMLE's were constructed using locally least favorable submodels as defined in van der Laan and Gruber, 2016, which are indexed by a multidimensional fluctuation parameter.  In this paper we define a canonical least favorable submodel in terms of a single dimensional epsilon for a $d$-dimensional parameter of interest.  One can view the clfm as the iterative analog to the one-step TMLE as constructed in van der Laan and Gruber, 2016.  It is currently implemented in several software packages we provide in the last section.  Using a single epsilon for the targeting step in TMLE could be useful for high dimensional parameters, where using a fluctuation parameter of the same dimension as the parameter of interest could suffer the consequences of curse of dimensionality.  The clfm also enables placing the so-called clever covariate denominator as an inverse weight in an offset intercept model.  It has been shown that such weighting mitigates the effect of large inverse weights sometimes caused by near positivity violations \parencite{Robins2007}.   

\end{abstract}
\end{titlepage}

\newpage
\section*{Introduction}
We offer a new way to construct a targeted maximum likelihood estimator for multidimensional parameters via defining the canonical least favorable submodel (clfm).   TMLE is a plug-in estimator so it follows that we might prefer to use the same model estimate for all dimensions of a parameter of interest.  The obvious example of such is a survival curve, in order to insure monotonicity of the estimates in time.  The clfm leads naturally to the construction of the one-step TMLE \parencite{Laan:2015ab}.  The resulting TMLE algorithm can be seen as an iterative version of the one-step TMLE in that both TMLE's use a single dimensional submodel in their construction. \\

The TMLE defined here-in can converge much faster than its one-step recursive counterpart when evaluating the efficient influence curve has a cost.  This is due to relatively few logistic regression fits as compared to very small recursions.  The procedure also enables placing the denominator of the clever covariate as an inverse weight in an offset intercept model, shown to stabilize large weights caused by near positivity violations.  In addition, like the one-step TMLE, the TMLE based on a clfm involves the use of a one-dimensional submodel, which avoids high dimensional regressions to perform the targeting step in the algorithm. \\ 

In this paper we will first review the TMLE basics and then construct the TMLE based on the clfm, giving an algorithm for its implementation, currently available in several R packages where simultaneous estimation is an option.  

\section{TMLE, a Brief Review}

 We refer the reader to Targeted Learning Appendix \parencite{Laan:2011aa} as well as \parencite{Laan:2015aa,Laan:2015ab, Laan:2006aa} for a more detailed look at the theory of TMLE.  Here we review the basics for the convenience of the reader. \\
 
Consider observed data, $O\sim P \in \mathcal{M}$, non-parametric, and a d-dimensional pathwise differentiable \parencite{Vaart:2000aa} parameter mapping, $\Psi:\mathcal{M}\longrightarrow \mathbb{R}^d$.   Consider our sample are iid copies from $P$.  The efficient influence curve or canonical gradient, $$D^{\star}_{\Psi}(P)(O)=(D^{\star}_{\Psi_1}(P)(O),...,D^{\star}_{\Psi_d}(P)(O))$$ is a d-dimensional function of the observed data $O$ and defined in terms of the distribution, $P$.  Its variance gives the generalized Cramer-Rao lower bound for the variance of any regular asymptotically linear estimator of $\Psi(P)$ \parencite{Vaart:2000aa}. \\
 
We will employ the notation, $P_{n}f(O)$ to be the empirical average of function, $f(\cdot)$, and $Pf(O)$ to be $\mathbb{E}_{P}f(O)$.  Define a loss function, $L(P)(O)$, which is a function of the observed data, O, and indexed at the distribution on which it is defined, $P$, such that $E_{P_0} L(P)(O)$ is minimized at the true observed data distribution, $P=P_0$. The TMLE procedure maps an initial estimate, $P_{n}^{0}\in \mathcal{M}$, of the true data generating distribution to $P_{n}^{\star}\in \mathcal{M}$ such that $P_{n}L(P_{n}^{\star})\leq P_{n}L(P_{n}^{0})$ and such that $P_{n}D^{\star}(P_{n}^{\star})=0_{d\times1}$. $P_{n}^{\star}$ is called the TMLE of the initial estimate $P_{n}^{0}$.  We can then write a second order expansion, $\Psi(P_{n}^{\star})-\Psi(P_{0})=(P_{n}-P_{0})D^{\star}(P_{n}^{\star})+R_{2}(P_{n}^{\star},P_{0})$. 

\subsection{Conditions for Asymptotic Efficiency}
Define the norm $\Vert f \Vert_{L^{2}(P)} = \sqrt{\mathbb{E}_{P}f^{2}}$. Assume the following TMLE conditions:

\begin{enumerate}
\item
$D^{\star}_{\Psi_j}(P_{n}^{\star})$ is in a P-Donsker class for all $j$. This condition can be dropped in the case of using CV-TMLE \parencite{Zheng:2010aa}.  

\item
$R_{2,j}(P_n^*,P_0)$ is $o_{p}(1/\sqrt{n})$ for all $j$.
\item
$D^{\star}_{\Psi_j}(P_{n}^{\star})\overset{L^{2}(P_{0})}{\longrightarrow} D^{\star}_{\Psi_j}(P_{0})$ for all $j$. 

\end{enumerate}

then $\sqrt n(\Psi(P_{n}^{\star})-\Psi(P_{0})) \overset{D}{\implies} N[0_{2\times1}, cov_{P_0}(D^{\star}_{\Psi}(P_{0})_{2\times2}]$ where 
$cov_{P_0}(D^{\star}_{\Psi}(P_{0})(O)$ is a $2\times2$ matrix in our case with the $(i,j)$ entry given as $E_{P_0} D^*_{\Psi_i}(P_0)(O)D^*_{\Psi_j}(P_0)(O)$.  The $i^{th}$ diagonal of $cov_{P_0}(D^{\star}_{\Psi}(P_{0})(O)$ is the variance of the $D^*_{\Psi_i}(P_0)$ and the limiting variance of $\sqrt{n}(\Psi_i(P_n^*) - \Psi_i(P_0))$ under TMLE conditions.  Thus, our plug-in TMLE estimates and CI's given by 

$$\Psi_{j}(P_{n}^{\star})\pm z_{\alpha}*\frac{\widehat{\sigma}_n(D_{j}^{\star}(P_{n}^{\star}))}{\sqrt{n}}$$ 

will be as small as possible for any regular asymptotically linear estimator at significance level, $1-\alpha$, where $Pr(\vert Z \vert \leq z_{\alpha})=\alpha$ for Z standard normal and $\widehat{\sigma}_n(D_{j}^{\star}(P_{n}^{\star}))$ is the sample standard deviation of $\{D_{j}^{\star}(P_{n}^{\star})(O_i) \mid i \in 1:n \}$ \parencite{Laan:2006aa}.  

\section{Mapping $P_{n}^{0}$ to $P_{n}^{\star}$: The Targeting Step}
The preceding section sketched the framework by which TMLE provides asymptotically efficient estimators for nonparametric models. Here we will explain how TMLE maps an initial estimate $P_{n}^{0}$ to $P_{n}^{\star}$, otherwise known as the targeting step. $P_{n}^{0}$ is considered to be the initial estimate for the true distribution, $P_{0}$. 

\begin{defn}
We can define a canonical 1-dimensional locally least favorable submodel (clfm) of an estimate, $P_{n}^{0}$, of the true distribution as
\begin{equation}
\{P_{n, \epsilon}^0 \text{ s.t } \frac{d}{d\epsilon}P_{n}L(P_{n, \epsilon}^0)\biggr\vert_{\epsilon=0}=\Vert P_{n} D^{\star}(P_{n}^{0})\Vert_{2}, \epsilon \in [-\delta,\delta]\} 
\end{equation}

where $P_{n, \epsilon}^0 = P_n^0$ and $\Vert \cdot \Vert_{2}$ is the euclidean norm.  We consider a $d-dimensional$ parameter mapping $\Psi:\mathcal{M}\longrightarrow \mathbb{R}^{d}$.
\end{defn} 

This definition only slightly differs slightly from the locally least favorable submodel (lfm) defined by Mark van der Laan \parencite{Laan:2015ab} in that we can define a clfm with only a single epsilon and in so far as the lfm is defined so the score with respect to the loss spans the efficient influence curve.  

\begin{defn}
A Universal Least Favorable Submodel (ulfm) of $P_{n}^{0}$ satisfies
\[
\frac{d}{d\epsilon}P_{n}L(P_{n}^{\epsilon})=\Vert P_{n}D^{\star}(P_{n}^{\epsilon})\Vert_{2}\text{ }\forall\epsilon\in(-\delta,\delta)
\]
and naturally, $P_n^{\epsilon = 0} = P_n^0$.  
\end{defn}

We can construct the universal least favorable submodel (ulfm) in terms of the clfm if we use the difference equation $P_{n}(L(P_{n,dt}^{0})-L(P_{n}^{0}))  \approx  \Vert P_{n}D^{\star}(P_{n}^{0})\Vert_{2}dt$, where $P_{n}^{dt} = P_{n, dt}^{0}$ is an element of the clfm of $P_n^0$. More generally, we can map any partition $t=m\times dt$ for an arbitrarily small, $dt$, to an equation $P_n(L(P_{n}^{t+dt})-L(P_{n}^{t}))  \approx  \Vert P_{n}D^{\star}(P_{n}^{t})\Vert_{2}dt$, where $P_{n}^{t+dt}$ is an element of the clfm of $P_{n}^{t}$. We therefore can recursively define the integral equation: $P_{n}(L(P_{n}^{\epsilon})-L(P_{n}^{0}))=\int_{0}^{\epsilon}\Vert P_{n}D^{\star}(P_{n}^{t})\Vert_{2}dt$ and  $P_n^\epsilon$ will thusly be an element of the ulfm of $P_n^0$.  For log likelihood loss, which is valid for both continuous outcome scaled between 0 and 1 as well as binary outcomes, an analytic formula for a ulfm of distribution with density, $p$, is therefore defined by the density $p_\epsilon=p\times exp(\int_0^\epsilon \Vert D^*(P^t) \Vert_2 dt)$ \parencite{Laan:2015ab} where $P^{t+dt}$ is an element of the clfm of $P^{t}$. \\

In applying the one-step TMLE, when the empirical loss is minimized at a given $\epsilon$, we will have solved, $\Vert P_{n}D^{\star}(P_{n}^{\epsilon})\Vert_{2}=0$. Therefore, the loss is decreased and all influence curve equations are solved simultaneously with a single $\epsilon$ in one step. Specifically, $P_{n}D^{\star}_{j}(P_{n}^{\star})=0$ for all $j$. Thus $P_{n}^{\star}=P_{n}^{\epsilon}$ and we have defined the required TMLE mapping.  

\subsection{The Iterative Approach Offered in This Paper}
With an iterative approach, we first find $P_{n, \epsilon_0}^0 = P_{n}^{1}$, that is an element of the clfm of $P_n^0$ such that
\begin{equation}
\frac{d}{d\epsilon}P_{n}L(P_{n, \epsilon}^{0})\biggr\vert_{\epsilon = \epsilon_0}=0
\end{equation} 

This initializes an iterative process where by 

\begin{equation}
\frac{d}{d\epsilon}P_{n}L(P_{n, \epsilon}^{j-1})\biggr\vert_{\epsilon = \epsilon_j}=0.
\end{equation} 

where $P_{n, \epsilon}^{j}$ is an element of the clfm of $P_{n}^{j-1}$.  When $\epsilon_j=0$, we stop the process and our TMLE is $P_{n}^{\star}=P_{n}^{j-1}$.

\subsection{CLFM Construction for Generalized Scenario}

Assume we have a parameter mapping as defined in the previous section, where the data is of the form $O = (W,A,Y)\sim P_0$ where $Y$ and $A$ are binary and $W$ is a vector of confounders.  We consider the likelihood factored according to $p_0(w,a,y) = \bar{Q}_0(a,w)^Y(1-\bar{Q}_0(a,w))^{1-Y}g_0(a \mid w)q_{W,0}(w)$.  We also assume we have efficient inflluence curve for the $jth$ component of the parameter of the form:

\begin{scriptsize}
\[
D_j^*(P_0)(O) = H_{1,j}(p_0)(A,W)(Y - \bar{Q}_0(A,W)) +  H_{2,j}(p_0)(A,W)(A - g_0(A,W))+H_{3,j}(A,W)(f(P_0)_j(A,W) - \Psi(P_0))
\]
\end{scriptsize}

where $\Psi(P_0) = E_0[H_{2,j}(O_i(f(P_0)_j(O)]$ and $E_0[H_{j,2}(O_i)=1$ for fixed function $H_{j,2}$.  Also note the dependence of the function $H_{1,j}(p_0)$ and $H_{2,j}(p_0)$ on the distribution. Now assume we have an initial estimate of $P_n^0$, of $P_0$, via an estimate, $p_n^0$, of the density $p_0$.  We define $p_n^0$ by estimates of factors of the likelihood.  That is, $\bar{Q}_n^0\approx \bar{Q}_0$, $g_n\approx g_0$, and $Q_{W,n}$ places a $q_{W,n}=1/n$ weight on every observation.  The latter is used to approximate the true distribution of $W$, $Q_{W,0}$.  A clfm of $P_n^0$ is defined by leaving $Q_{W,n}$ fixed and defining 
\[
\bar{Q}_{n,\epsilon}^0(A,W) = expit\left(logit(\bar{Q}_n^0(A,W)) + \epsilon \biggr\langle H_1(P_n^0)(A,W), \frac{P_n D^*(P_n^0)}{\Vert P_n D^*(P_n^0) \Vert_2}\biggr\rangle_2\right)
\]
and 
\[
g_{n,\epsilon}^0(A\mid W) = expit\left(logit(g_n^0(A \mid W)) + \epsilon \biggr\langle H_2(P_n^0)(A,W), \frac{P_n D^*(P_n^0)}{\Vert P_n D^*(P_n^0) \Vert_2}\biggr\rangle_2\right)
\]

 where $\Vert \cdot \Vert_2$ is the euclidean norm induced by dot product, $\langle \cdot, \cdot \rangle$.  In the usual case we have
$P_n H_{2,j}(f(P_n^0)_{j} - \Psi(P_n^0))=0$ and therefore $p_{n,\epsilon}^0$ defines an element, $P_{n,\epsilon}^0$, of a clfm of $P_n^0$. 

\subsection{General TMLE Algorithm using the clfm for Point Treatment Parameters}

\textbf{Initialization} \\
We start the iterative process with our initial estimate $p_{n}^{0}$ as defined in the previous subsection. 

\begin{align*}
P_{n}L(P_{n}^{0}) &= -\frac{1}{n}\sum_{i=1}^{n}\left[Y_{i}\text{log}\bar{Q}_{n}^{0}(A_{i},W_{i})+(1-Y_{i})\text{log}(1-\bar{Q}_{n}^{0}(A_{i},W_{i}))\right]\\
& -\frac{1}{n}\sum_{i=1}^{n}\left[A_{i}\text{log} g_{n}^{0}(A_{i} \mid W_{i})+(1-A_{i})\text{log}(1-g_{n}^{0}(A_{i} \mid W_{i}))\right]\\
&=L_{0}\text{ our starting loss}\\
\end{align*}

\subsubsection*{The Targeting Step}
Starting with $m=0$

\textbf{step 2:}

Compute $H_1(P_n^m)(A,W)$, $H_2(P_n^m)(A,W)$ and  $H_{2}(A,W)$ over the data and then check the following:  
If $\vert P_{n}D^{\star}_{j}(P_{n}^{m})\vert< \hat{\sigma}(D^{\star}_j(P_n^m))/n$ for all $j$ then $P_{n}^{\star}=P_{n}^{m}$ and go to step 4. This insures that we stop the process once the bias is second order. Note, $\hat{\sigma}(\cdot)$ refers to the sample standard deviation operator. Otherwise set $m = m+1$ and go to step 3. 

\textbf{step 3}
We perform a pooled logistic regression with $Y$ as the outcome, \\ offset = $logit(\bar{Q}_{n}^{m-1})(A,W)$ and so-called clever covariate, 
$$\biggr\langle(H_1(P_n^{m-1})(A,W),\frac{P_{n} D(P_{n}^{m-1})}{\Vert P_{n}D(P_{n}^{m-1})\Vert_{2}}\biggr\rangle_{2}.$$

and $A$ as the outcome, offset $logit(g_{n}^{m-1})(A\mid W)$ and so-called clever covariate, 
$$\biggr\langle(H_2(P_n^{m-1})(A,W),\frac{P_{n} D(P_{n}^{m-1})}{\Vert P_{n}D(P_{n}^{m-1})\Vert_{2}}\biggr\rangle_{2}.$$

Assume $\epsilon_j$ is the coefficient computed from the above pooled regression.  We then update the models as per below, using euclidean inner product notation, $\langle \cdot, \cdot \rangle_{2}$:
{\footnotesize{}
\begin{equation}
\bar{Q}_{n}^{m}=expit\left(logit(\bar{Q}_{n}^{m-1})-\epsilon_j\biggr\langle(H_1(P_n^{m-1})(A,W),\frac{P_{n} D(P_{n}^{m-1})}{\Vert P_{n}D(P_{n}^{m-1})\Vert_{2}}\biggr\rangle_{2}\right)
\end{equation}

and

\begin{equation}
g_{n}^{m}(A\mid W)=expit\left(logit(g^{m-1}(A\mid W))-\epsilon_j\biggr\langle(H_2(P_n^{m-1})(A,W),\frac{P_{n} D(P_{n}^{m-1})}{\Vert P_{n}D(P_{n}^{m-1})\Vert_{2}}\biggr\rangle_{2}\right)
\end{equation}}

\subsubsection*{Possible alternative targeting step to ameliorate near positivity violations}
We can alternatively perform a pooled logistic regression as follows.  For all observations we use $Y$ as the outcome, offset = $logit(\bar{Q}_{n}^{m-1})(A,W)$.  We denote the denominator of $H_{1,j}(P_n^{m-1})$ as $g_j(P_n^{m-1})$, which, in some cases is a fixed propensity score, $g(P_n^{m-1})$.  We can use its inverse as a weight in a logistic regression model with covariate 

$$g(P_n^{m-1})(A\mid W)^{-1}\biggr\langle(H_1(P_n^{m-1})(A,W),\frac{P_{n} D(P_{n}^{m-1})}{\Vert P_{n}D(P_{n}^{m-1})\Vert_{2}}\biggr\rangle_{2}.$$  

We then stack all observations using $A$ as the outcome, offset, $logit(g_{n}^{m-1})(A\mid W)$ and so-called clever covariate, 

$$\biggr\langle(H_2(P_n^{m-1})(A,W),\frac{P_{n} D(P_{n}^{m-1})}{\Vert P_{n}D(P_{n}^{m-1})\Vert_{2}}\biggr\rangle_{2}.$$  

Thus we use a weight of 1 for when $A$ is the outcome because $H_2(P_n^{m-1})(A,W)$ generally does not have large values.  We then update the models similarly as before upon solving for the coefficient $\epsilon_j$.  With either regression scheme we solve the same score equation so either are appropriate for the targeting step.\\

Once we are done with the targeting step we define the distribution, $P_{n}^{m}$, via factors of the density for the outcome model and propensity score, i.e., $\bar{Q}_{n}^{m}(A,W)$ and $g_{n}^{m}(A\vert W)$, while placing a weight of $1/n$ for each observation as an estimate of the true distribution of $W$.  Return to step 2.  

\textbf{step 4}

Our estimate is $\hat{\Psi}(P_{n})=\Psi(P_{n}^{\star})$ which is really only dependent on $\bar{Q}_{n}^{\star}$ and the empirical distribution.

\subsubsection{R Software employing the clfm}
Currently there are three packages which employ the iterative TMLE as presented in this paper for parameters with influence curves of the form as in this paper.  Note to the reader, we have yet to implement the weighted intercept targeting scheme as discussed in step 3 of the algorithm in section 4.  
\begin{itemize}
\item
tmle3, https://github.com/tlverse/tmle3 \parencite{tmle3}

There are various parameters for which one can perform a TMLE estimator, including variable importance measure for continuous variables \parencite{shift}, treatment effect among the treated, causal risk difference, treatment specific mean and more.  

\item
gentmle2, https://github.com/jeremyrcoyle/gentmle2 \parencite{gentmle2}
The reader may note this clfm is what is employed in this R package when specifying the approach as "line". An lfm with epsilon the same dimension as the parameter is employed with the "full" option.   Other than causal risk difference and treatment specific mean, there is also the variance of treatment effect \parencite{catesurvival} as well as the mean under stochastic intervention \parencite{Munoz-ID:2012aa}.   
\item
cateSurvival, https://github.com/jlstiles/cateSurvival \parencite{cateSurvival}

This package implements a TMLE estimator for $\Psi_{k,t}(P) = \int k\left(\frac{x- t}{h} \right) \mathbb{E}_P\mathbb{I}(B(W) > x) dx$ which is kernel-smoothed version of the non-pathwise differentiable parameter, $\mathbb{E}_P\mathbb{I}(B(W) > t)$, where $B(W)$ is the treatment effect function or TE function, defined by $\mathbb{E}_P[Y \mid A=1, W] - \mathbb{E}_P[Y \mid A=0, W]$.  The non-pathwise differentiable parameter gives the probability a subject selected at random will have treatment effect beyond the level $t$.  It can be thought of as a "survival" of the treatment effect function because it is monotonically decreasing.  It is also more familiarly, 1 - CDF of the random variable that gives the treatment effect for a subject drawn at random.  The user can select the kernel according to its support and its order.  \\

\end{itemize}

\newpage
\printbibliography
\clearpage

\end{document}